\documentclass[preprint,showpacs,preprintnumbers,amsmath,amssymb]{revtex4}

\usepackage{graphicx}% Include figure files
\usepackage{dcolumn}% Align table columns on decimal point
\usepackage{bm}% bold math 

\begin{document}

\title{ Effect of isospin dependent cross-section on nuclear stopping.\\}
% Force line breaks with \\

\author{Anupriya Jain}
\author{Suneel Kumar}%
\email{suneel.kumar@thapar.edu}

\affiliation{% 
School of Physics and Materials Science, Thapar University, Patiala-147004, Punjab (India)\\
}%

\date{\today}% It is always \today, today,




%%%%%%%%%%%%%%%%%%%%%%%%%%%%%%%%%%%%%%%%%%%%%%%%%%%%%%%%%%%%%%%%%%%%%%%%%%%%%%%%%%%%%%%%%%%%%

\maketitle
\section{Introduction}

Heavy ion collisions (HIC) provide the unique opportunity to produce small amount of nuclear matter with high density and high temperature in laboratory. By measuring the final products of the collisions, it is possible to learn about the fundamental properties of hot and compressed nuclear matter, namely the nuclear equation of state (EOS), which is of great importance in our understanding of the evolution of macroscopic objects like neutron star and supernova \cite{1}. The colliding nuclei, however, are finite size objects and thus many dynamical effects cannot be ignored. For this reason, in order to link the experimental observations and the EOS in HIC, we need transport model in which the nucleon-nucleon two body collisions and the mean field effects are carefully treated.\\ 

Nuclear stopping in heavy ion collisions (HIC) has been studied by means of rapidity distribution and asymmetry of nucleon momentum distribution. It is an important quantity in determining the outcome of a reaction. Fen Fu {\it et al.,} \cite{2} calculate both the radial flow and the degree of nuclear stopping in Pb + Pb and Ni + Ni at 0.4, 0.8 and 1.2 GeV/neucleon. They found that the expansion velocity as well as the degree of nuclear stopping are higher in the heavier system at all energies.\\
The stopping parameter can be defined as \cite{3}:
\begin{equation}
Q_{ZZ}=\sum_{i}2p_z^2(i)-p_x^2(i)-p_y^2(i)
\end{equation}

\begin{equation}
R=\frac{2}{\pi}\frac{\left(\sum_{i}p_{\perp}(i)\right)}{\left(\sum_{i}p_{\parallel}(i)\right)}
\end{equation}
In this study our aim is to pin down the influence of different cross-sections (isospin dependent or isospin independent) on nuclear stopping.\\

\section{ISOSPIN-dependent QUANTUM MOLECULAR DYNAMICS (IQMD) MODEL}

Calculations are carried out within the framework of Isospin dependent Quantum Molecular Dynamics (IQMD) \cite{4} model, which is a modified version of QMD \cite{5} model. The IQMD is a semi-classical model
which describes the heavy-ion collisions on an event by event basis. For more details, see ref.\cite{4}.\\
In IQMD model, the centroid of each nucleon propagates under the classical equations of motion.
\begin{equation}
\frac{d\vec{r_i}}{dt}~=~\frac{d\it{H}}{d\vec{p_i}}~~;~~\frac{d\vec{p_i}}{dt}~=~-\frac{d\it{H}}
{d\vec{r_i}}~~\cdot
\end{equation}
The $H$ referring to the Hamiltonian reads as:
\begin{equation}
~H~=~\sum_{i}\frac{p_i^2}{2m_i}+V^{ij}_{Yukawa}+V^{ij}_{Coul}+V^{ij}_{skyrme}+
V^{ij}_{sym}\cdot
\end{equation}

During the propagation, two nucleons are
supposed to suffer a binary collision if the distance between their centroids
\begin{equation}
|r_i-r_j| \le \sqrt{\frac{\sigma_{tot}}{\pi}}, \sigma_{tot} = \sigma(\sqrt{s}, type),
\end{equation}\\
%%%%%%%%%%%%%%%%%%%%%%%%%%%%%%%%%%%%%%%%%%%%%%%%%%%%%%%%%%%%%%%%%%%%%%%%%%%%%%%%%%%%%%%%%%%%%
{\large{\bf Results and Discussion}}\\

For the present analysis simulations are carried out for several  thousand events at the incident energy of 100 MeV/nucleon for the systems $^{124}Ag_{47}+^{124}Ag_{47}$, $^{124}Cd_{48}+^{124}Cd_{48}$, $^{124}In_{49}+^{124}In_{49}$, $^{124}Sn_{50}+^{124}Sn_{50}$, $^{124}I_{53}+^{124}I_{53}$, $^{124}Cs_{55}+^{124}Cs_{55}$, $^{124}Ba_{56}+^{124}Ba_{56}$, $^{124}Pr_{59}+^{124}Pr_{59}$.
\begin{figure}
\hspace{-2.0cm}\includegraphics[scale=0.50]{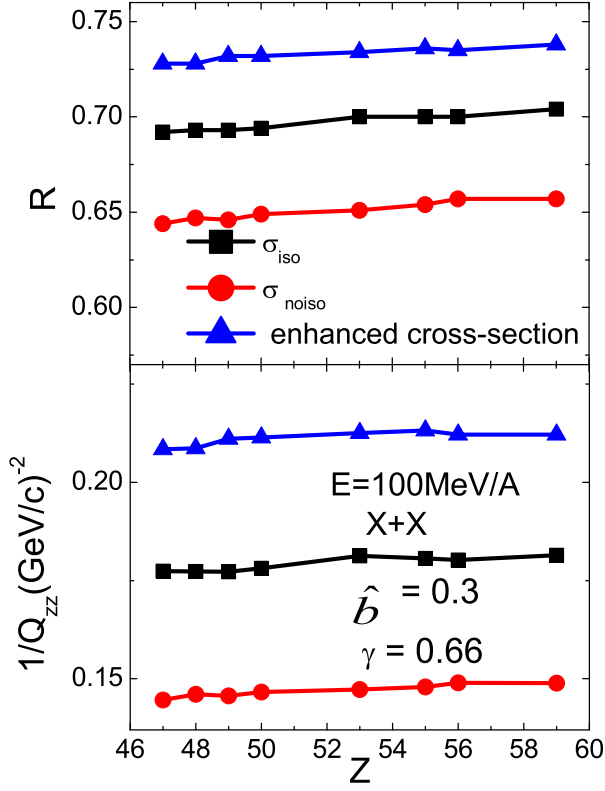}
\caption{\label{fig:1} Variation of $1/Q_{ZZ}$ and R with Z for X= $^{124}Ag_{47}$, $^{124}Cd_{48}$, $^{124}In_{49}$, $^{124}Sn_{50}$, $^{124}I_{53}$, $^{124}Cs_{55}$, $^{124}Ba_{56}$, $^{124}Pr_{59}$}.
\end{figure}
In fig.1, we display the variation of stopping parameters e.g. $1/Q_{ZZ}$ and R as a function of Z, for the symmetry energy of $32(\rho/\rho_o)^{\gamma}$~ MeV for ~${\gamma}=0.66$ for three different cross-sections $\sigma_{iso}$, $\sigma_{noiso}$ (isospin dependent and isospin independent) and enhanced cross-section. We observe that R and $1/Q_{ZZ}$ behave in a similar way. Maximum stopping is observed in case of system with lesser neutron content. Moreover, in the systems with more neutron content, the role of symmetry energy could be larger, whereas the effects caused by an isospin-dependent cross-section could play a dominant role in systems with less neutron content. We observe the combined effect of symmetry energy and the cross-section, because for isospin dependent cross-section a neutron-neutron or a proton-proton cross-section is a factor of three lower than the neutron-proton cross-section. Due to which the attractive forces between the neutron-proton increases which enhance the number of collision and hence stopping increases than isospin independent cross-section this observation is in agreement with the observation in ref. \cite{6}. Moreover, in third cross-section we enhance the cross-section of $\sigma_{nn}$ and  $\sigma_{pp}$ due to which value of stopping become maximum. Moreover, from fig.1 we can see that value of R for Sn+Sn is 0.7022, this observation is in agreement with the observation in ref. \cite{7}.\\

%%%%%%%%%%%%%%%%%%%%%%%%%%%%%%%%%%%%%%%%%%%%%%%%%%%%%%%%%%%%%%%%%%%%%%%%%%%%%%%%%%%%%%%%%%%%%%%%%%%%%%%%%%%%%%%%5

\end{document}